\documentclass[sigconf,nonacm,10pt,table,screen]{acmart}
\usepackage{alphaledger_arxiv}
\usepackage[T1]{fontenc}
\usepackage[utf8]{inputenc}
\usepackage{microtype}
\usepackage{graphicx}
\usepackage{booktabs}
\usepackage{multirow}
\usepackage{amsmath,amsfonts}
\usepackage{xcolor}
\usepackage{colortbl}
\definecolor{oursrow}{RGB}{235,243,249}
\usepackage{pifont}
\usepackage{tabularx,array}

\title[AlphaLedger]{AlphaLedger: Portfolio-Grounded On-Policy Distillation for Regime-Aware Alpha Mining}
\author{Xiang Li}
\email{xli906@connect.hkust-gz.edu.cn}
\authornote{Equal contribution.}
\affiliation{%
  \institution{The Hong Kong University of Science and Technology (Guangzhou)}
  \city{Guangzhou}
  \country{China}}

\author{Zikai Wei}
\email{weizikai@idea.edu.cn}
\authornotemark[1]
\affiliation{%
  \institution{International Digital Economy Academy}
  \city{Shenzhen}
  \country{China}}

\author{Yiyan Qi}
\email{qiyiyan@idea.edu.cn}
\authornotemark[1]
\affiliation{%
  \institution{International Digital Economy Academy}
  \city{Shenzhen}
  \country{China}}

\author{Wanyun Zhou}
\email{wzhou266@connect.hkust-gz.edu.cn}
\affiliation{%
  \institution{The Hong Kong University of Science and Technology (Guangzhou)}
  \city{Guangzhou}
  \country{China}}

\author{Xiang Liu}
\email{xliu886@connect.hkust-gz.edu.cn}
\affiliation{%
  \institution{The Hong Kong University of Science and Technology (Guangzhou)}
  \city{Guangzhou}
  \country{China}}

\author{Penglei Sun}
\email{psun012@connect.hkust-gz.edu.cn}
\affiliation{%
  \institution{The Hong Kong University of Science and Technology (Guangzhou)}
  \city{Guangzhou}
  \country{China}}

\author{Zhenheng Tang}
\email{zhtang.ml@ust.hk}
\affiliation{%
  \institution{The Hong Kong University of Science and Technology (Guangzhou)}
  \city{Guangzhou}
  \country{China}}

\author{Yongqi Zhang}
\email{yongqizhang@hkust-gz.edu.cn}
\authornote{Corresponding authors.}
\affiliation{%
  \institution{The Hong Kong University of Science and Technology (Guangzhou)}
  \city{Guangzhou}
  \country{China}}

\author{Jian Guo}
\email{guojian@idea.edu.cn}
\authornotemark[2]
\affiliation{%
  \institution{International Digital Economy Academy}
  \city{Shenzhen}
  \country{China}}

\author{Xiaowen Chu}
\email{xwchu@hkust-gz.edu.cn}
\authornotemark[2]
\affiliation{%
  \institution{The Hong Kong University of Science and Technology (Guangzhou)}
  \city{Guangzhou}
  \country{China}}

\renewcommand{\shortauthors}{Xiang Li et al.}

\begin{document}

\begin{abstract}
Hosted LLMs enable automated alpha mining but incur recurring API costs and expose research context to external providers. Training a local generator requires supervision aligned with portfolio utility: a predictive factor may add little after redundancy and trading costs. This incremental value is assessed only after the factor program is complete, whereas generation proceeds sequentially. Deployment must also adapt to evolving stock-specific evidence and changes in factor effectiveness across market regimes. We propose \textbf{AlphaLedger} to address these challenges. Economic-context on-policy distillation (EC-OPD) conditions a frozen teacher on completed factors and their cost-adjusted marginal portfolio contributions, then distills its legal-action distributions along the student's own trajectories. The deployed generator requires neither teacher calls nor outcome inputs. EviPool constructs a dynamic stock universe from lagged multi-source evidence, while a regime-residual gate (RRG) reweights factors using matured, state-matched returns with shrinkage and bounded deviations from equal weighting. On Chinese A-shares, AlphaLedger achieves a \textbf{55.4\%} annualized net return, a Sharpe ratio of \textbf{2.06}, and a maximum drawdown of \textbf{10.8\%}. In a common CSI500 universe, its annualized net return exceeds the strongest baseline by 43.6\% (31.6\% vs.\ 22.0\%), with a 13.5\% higher Sharpe ratio.
\end{abstract}

\maketitle

\section{Introduction}

Alpha factor mining constructs signals for predicting future stock returns \cite{luo2026alphabench}. Formulaic factors are executable programs that transform market observations into cross-sectional stock scores, supporting interpretable analysis and systematic portfolio construction \cite{yu2026towards,luo2025efs}. Automated discovery has progressed from reinforcement learning and structured search \citep{yu2023alphagen,chen2026alphasage,yang2026alphacfg} to LLMs that translate economic hypotheses into programs \citep{li2024fama,tang2025alphaagent}. Agentic systems further coordinate hypothesis generation, implementation, backtesting, and refinement \citep{li2025rdagentq,han2026quantaalpha,wang2026factorminer}.

Many systems access LLMs through external APIs and place evaluation feedback in subsequent prompts or experience memory \cite{shahout2026fast,patil2024gorilla}. Repeated calls incur inference costs and send research context to external providers. Post-training local generators offer an alternative by incorporating experience into model parameters \citep{zhang2026quantevolver,wang2026alphadiverse}. However, a factor that predicts returns well in isolation may duplicate existing signals or increase trading costs. Learning useful factors therefore requires portfolio-aware evaluation, while deploying them requires adapting their eligible stocks and weights. This raises two challenges.

\textbf{Ch1: Aligning generation with incremental portfolio value.}
A candidate's economic value depends on its contribution to an existing factor ensemble \citep{yu2023generating}, with trading costs affecting the resulting profitability \citep{novy2016taxonomy}. This contribution is measured by evaluating the completed program, whereas generation proceeds through choices of operators, fields, and windows. A terminal scalar reward can train the generator, %but does not by itself specify which intermediate actions should be preferred 
but does not directly provide preferences among intermediate actions
\citep{lightman2024let}. The challenge is to translate portfolio-level evidence into action-level supervision that guides the generator toward factors with greater incremental value after costs.
% A candidate's economic value depends on its contribution to an existing factor ensemble after costs. This value can be evaluated only once the program is complete, whereas generation proceeds through choices of operators, fields, and windows. A scalar reward can train the generator but does not directly specify preferences among intermediate actions. The challenge is to translate portfolio-level evidence into action-level supervision that guides the generator toward factors with greater incremental value after costs.
% The challenge is to turn portfolio evidence into informative generation supervision while keeping evaluation outcomes out of the deployed generator's inputs.

\textbf{Ch2: Adapting deployment to evolving evidence.}
Index universes provide standard benchmarks, but their membership rules do not directly incorporate newly available stock-specific evidence, so relevant opportunities may emerge
outside the eligible set. Fixed factor weights likewise ignore changes in effectiveness across market regimes \citep{ang2012regime}. Adaptive weighting must also contend with limited observations within each regime, which make performance estimates noisy.

\begin{figure*}[t]
\centering
\includegraphics[width=\textwidth]{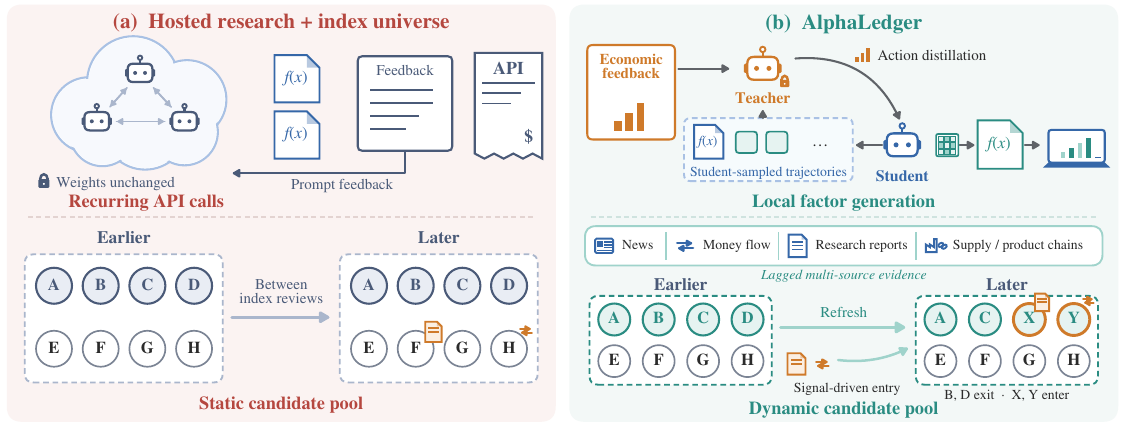}
\caption{API-based alpha mining and AlphaLedger. AlphaLedger learns a local generator from portfolio evidence and deploys its factors with an evidence-driven universe and adaptive weights.}
\label{fig:motivation}
\end{figure*}

% We propose \textbf{AlphaLedger}, which connects portfolio-grounded training to evidence-driven deployment (Figure~\ref{fig:motivation}). To address \textbf{Ch1}, \textbf{economic-context on-policy distillation (EC-OPD)} evaluates each student-generated factor by replacing a reference factor of the same economic intent and measuring its cost-adjusted portfolio contribution. A frozen teacher receives the completed program and its contribution evidence, then supplies distributions over legal next actions at the student's generation prefixes. The student distills these distributions along its own sampled trajectories \citep{wu2026lightning}; outcomes enter only the teacher's context. Deployment requires neither teacher calls nor outcome inputs to the generator.

We propose \textbf{AlphaLedger}, which combines portfolio-grounded training with evidence-driven deployment (Figure~\ref{fig:motivation}).
To address \textbf{Ch1}, \textbf{economic-context on-policy distillation (EC-OPD)} uses matured evaluation records to assess each student-generated factor's marginal contribution to an existing ensemble's after-cost portfolio performance.
Conditioned on the completed factor and its contribution evidence, a frozen teacher provides distributions over grammar-valid next actions at the student's generation prefixes.
The student distills these distributions along its own sampled trajectories \citep{wu2026lightning}, without receiving portfolio outcomes as inputs.
% At deployment, it generates factors locally without the teacher or evaluation feedback.

To address \textbf{Ch2}, \textbf{EviPool} constructs a dynamic candidate
stock universe from lagged multi-source evidence independently of factor scores. The \textbf{regime-residual gate (RRG)} adjusts factor weights using matured returns from matching market states, shrinking estimates toward aggregate performance and bounding deviations from equal weighting. 
% The generator and selected factor library remain fixed during evaluation; the universe and weights adapt.

On Chinese A-shares, the complete framework achieves a 55.4\% annualized net return, a Sharpe ratio of 2.06, and a maximum drawdown of 10.8\%. In a common CSI500 universe, AlphaLedger earns 31.6\% annually with a Sharpe ratio of 1.43, compared with 22.0\% and 1.26 for the strongest baseline. Our contributions are threefold:
\begin{itemize}\setlength{\itemsep}{1pt}\setlength{\parskip}{0pt}
\item We propose \textbf{EC-OPD}, which uses cost-adjusted marginal portfolio contributions as privileged teacher context for supervising a local generator on its own sampled trajectories.
\item We design \textbf{EviPool} and \textbf{RRG} to adapt stock eligibility and factor weights to market evidence under point-in-time constraints.
\item We evaluate AlphaLedger against seven baseline families under a shared trading protocol, with component ablations and repeated generations to assess the sources and sampling sensitivity of its performance.
\end{itemize}

\section{Related Work}
\label{sec:related}

\subsection{Automated Alpha Mining}
Automated alpha mining spans symbolic search, LLM research agents, and generator post-training. AlphaGen discovers factors that improve a combined predictor with reinforcement learning \citep{yu2023alphagen}, and AlphaSAGE and AlphaCFG use structure-aware generative flow networks and grammar-guided search \citep{chen2026alphasage,yang2026alphacfg}. Among LLM methods, FAMA introduces a neural-symbolic mining agent \citep{li2024fama}, Alpha-GPT supports human--AI collaboration \citep{wang2025alphagpt}, and AlphaAgent regularizes originality, hypothesis consistency, and complexity \citep{tang2025alphaagent}; R\&D-Agent-Quant coordinates factor--model optimization \citep{li2025rdagentq}, QuantaAlpha and CogAlpha evolve research trajectories and factor code \citep{han2026quantaalpha,liu2026cogalpha}, and FactorMiner maintains reusable skills and experience memory \citep{wang2026factorminer}. QuantEvolver applies reinforcement fine-tuning to factor generation \citep{zhang2026quantevolver}, and AlphaDiverse combines supervised fine-tuning with joint GRPO to train local research agents \citep{wang2026alphadiverse}. However, prompt-based refinement keeps experience outside model parameters, and scalar rewards do not provide action-level supervision informed by a factor's marginal contribution to net portfolio performance.

\subsection{Investment Universes}
Alpha-mining studies typically define candidates by index membership or market coverage: CSI300 and CSI500 in AlphaGen and AlphaForge \citep{yu2023alphagen,shi2025alphaforge}, CSI500 and the S\&P500 in AlphaAgent and AlphaMemo \citep{tang2025alphaagent,yu2026alphamemo}, and CSI1000 in FactorMiner \citep{wang2026factorminer}. AlphaEvolve uses 1,026 NASDAQ stocks filtered by data availability and price \citep{cui2021alphaevolve}, AlphaDiverse adds a broader A-share universe \citep{wang2026alphadiverse}, and QuantEvolve evaluates strategies on six U.S.\ stocks \citep{yun2025quantevolve}. These settings support comparison within common opportunity sets, but eligibility follows benchmark membership or data coverage; even when index constituents change, they are not selected on newly available stock-specific evidence, whose role in constructing the candidate universe remains largely unexamined.

\section{Method}
\label{sec:method}

\begin{figure*}[t]
\centering
\includegraphics[width=\textwidth]{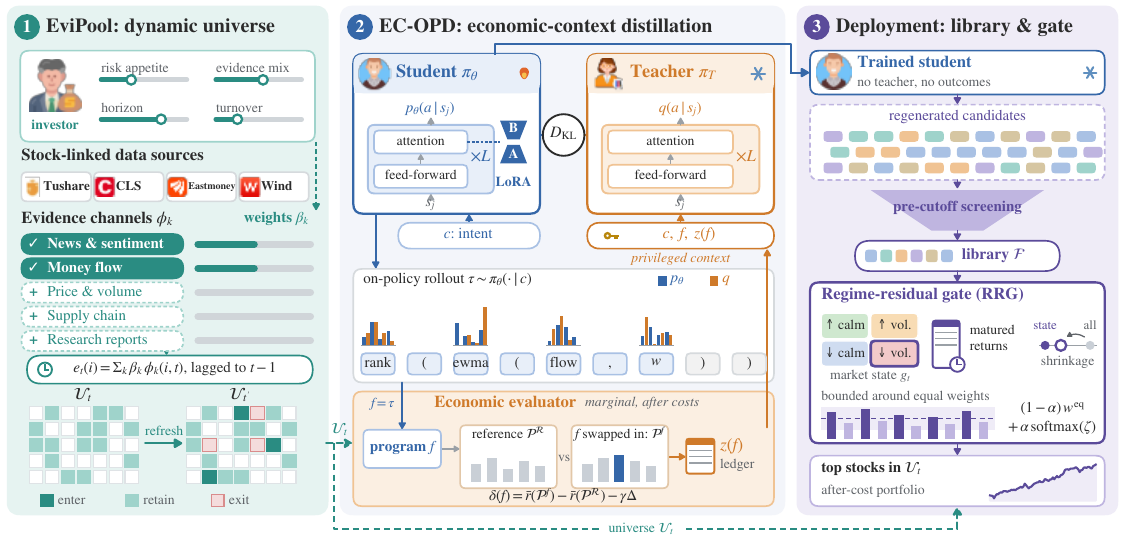}
\caption{Overview of AlphaLedger. \ding{172}~EviPool selects a stock universe from lagged evidence. \ding{173}~EC-OPD evaluates student programs and distills a frozen teacher conditioned on their economic feedback. \ding{174}~The trained student regenerates a fixed factor library, and RRG adapts its weights using matured returns. Checked evidence channels are evaluated; other inputs illustrate extensions.}
\label{fig:pipeline}
\end{figure*}

AlphaLedger separates three decisions (Figure~\ref{fig:pipeline}; information boundaries in Table~\ref{tab:boundaries}): \emph{where} factors trade (EviPool, \S\ref{sec:pool}), \emph{how} factors are generated (EC-OPD, \S\ref{sec:ecopd}), and \emph{how much} each factor is trusted (RRG, \S\ref{sec:rrg}), each using only information available at decision time.

\subsection{Problem Formulation}
\label{sec:setting}
Let $i$ index stocks and $t$ trading days. A factor $f$ is a typed-grammar program mapping market and news information available by close $t-1$ to a score $f(i,t)$; $\tilde f(i,t)\in[0,1]$ is its cross-sectional percentile. Given a factor library $\mathcal F$ and a candidate stock universe $\mathcal U_t$, the stock-ranking score is
\begin{equation}
S_t(i)=\textstyle\sum_{f\in\mathcal F}w_{f,t}\,\tilde f(i,t),
\label{eq:score}
\end{equation}
where $w_{f,t}\ge0$ are factor weights summing to one. The account selects high-scoring stocks within $\mathcal U_t$ under the trading rules in \S\ref{sec:setup}. We write $\mathrm{Top}_K(S;\mathcal U)$ for the $K$ highest-scoring stocks in $\mathcal U$ under score $S$.

A student generator $\pi_\theta$, with trainable parameters $\theta$, writes $f$ as a sequence $\tau=(a_1,\ldots,a_J)$ of $J$ grammar actions. During training, its context $c$ specifies an economic intent, its objective, and grammar instructions. At step $j$, the prefix $s_j=(c,a_{<j})$ contains the preceding actions, and $\mathcal A(s_j)$ is the legal next-action set. Training labels must mature by cutoff $t_0$; deployment uses only information available by $t-1$.

\subsection{EviPool: Evidence-Driven Universe}
\label{sec:pool}
EviPool determines membership in $\mathcal U_t$ from stock-linked evidence, leaving stock ranking to $S_t$ (\textbf{Ch2}).

\paragraph{Evidence score.} Evidence channel $k$ maps lagged observations to $\phi_k(i,t)\in[0,1]$. Fixed channel weights $\beta_k$ define the admission score
\begin{equation}
e_t(i)=\textstyle\sum_k\beta_k\phi_k(i,t),\quad
\beta_k\ge0,\quad\sum_k\beta_k=1.
\label{eq:evidence}
\end{equation}
Thus, $e_t$ selects the universe, whereas $S_t$ ranks stocks within it; $e_t$ is not added to $S_t$. The two stages may share raw inputs. The news channel $\phi_{\mathrm{news}}$ averages cross-sectional percentiles of recent issuer-linked news count and mean sentiment. The flow channel $\phi_{\mathrm{flow}}$ averages within-stock historical percentiles of five-day net money flow divided by circulating market value and of large-order imbalance.

\paragraph{Membership update.} Let $\mathcal E_t$ contain stocks passing price, listing-age, and liquidity filters, and let $\mathrm{rk}_t(i)$ rank $e_t(i)$ within $\mathcal E_t$ (highest score first). At scheduled refreshes, a pool of target size $N$ uses retention rank $M>N$:
\begin{equation}
\begin{aligned}
\mathcal I_t&=\{i\in\mathcal U_{t^-}\cap\mathcal E_t:\mathrm{rk}_t(i)\le M\},\\
\mathcal U_t&=\mathcal I_t\cup\mathrm{Fill}_t.
\end{aligned}
\label{eq:pool}
\end{equation}
Here $\mathcal U_{t^-}$ is the pool immediately before the update, $\mathcal I_t$ contains retained members, and $\mathrm{Fill}_t$ adds the highest-scoring eligible non-members until size $N$ is reached. Between refreshes, eligibility exits are removed and vacancies filled. The retention band limits membership churn.

\subsection{EC-OPD: Economic-Context On-Policy Distillation}
\label{sec:ecopd}
EC-OPD turns program-level economic feedback into teacher guidance on the student's own action prefixes (\textbf{Ch1}).

\paragraph{Economic credit.} Let $\mathcal R$ be a reference factor library with equal-weight score $S_t^{\mathcal R}$. Replacing a fixed same-intent reference slot with candidate $f$ gives $S_t^{\mathcal R\leftarrow f}$. Their respective $\mathrm{Top}_K$ portfolios are $\mathcal P_t^{\mathcal R}$ and $\mathcal P_t^f$. For entry date $t$ and holding horizon $h\in\mathcal H$, where $\mathcal H$ is the evaluated horizon set, define
\begin{equation}
\delta_t^h(f)=\tfrac1h\big[\bar r_t^{\,h}(\mathcal P_t^f)-\bar r_t^{\,h}(\mathcal P_t^{\mathcal R})-\gamma\Delta_t\big],
\label{eq:credit}
\end{equation}
where $\bar r_t^{\,h}$ is the mean stock return from close $t$ to close $t+h$, $\Delta_t=|\mathcal P_t^f\setminus\mathcal P_t^{\mathcal R}|/K$ is the replaced fraction, and $\gamma$ is a round-trip cost rate. Scores use close $t-1$, and labels exit by $t_0$. This dailyized replacement proxy is not a full-account P\&L difference. Let $\bar\delta(f)$ average credit over dates and horizons. We evaluate both $f$ and $-f$ on the same training labels. The ledger $z(f)$ reports the higher-credit direction, its mean and horizon/half summaries, and the opposite-direction mean. The teacher sees the original program with this direction-aware ledger; its token trajectory is unchanged. Reference factors are not deployed.

\paragraph{Outcome-conditioned teacher.} Both models start from pretrained base LLMs. The teacher $\pi_T$ is frozen; for each student rollout $\tau$ decoding to $f$, it receives the completed program and ledger $z(f)$ as privileged context. At prefix $s_j$, teacher and student distributions over legal action $a$ are
\begin{align}
q_j(a)&=\mathrm{softmax}_{a\in\mathcal A(s_j)}\,\ell_T(a\mid s_j,f,z(f)),\label{eq:teacher}\\
p_j^\theta(a)&=\mathrm{softmax}_{a\in\mathcal A(s_j)}\,\ell_\theta(a\mid s_j),\label{eq:student}
\end{align}
where $\ell_T$ and $\ell_\theta$ are next-action logits. The teacher uses the finished program's outcome to guide alternatives at each prefix; these preferences are not measured causal effects of individual actions.

\paragraph{On-policy distillation.} The student minimizes mean reverse KL divergence on its own sampled prefixes:
\begin{equation}
\mathcal L_{\mathrm{EC}}(\theta)=
\mathbb E_{c,\,\tau\sim\pi_\theta(\cdot\mid c)}
\Big[\frac1{|\mathcal J(\tau)|}\sum_{j\in\mathcal J(\tau)}D_{\mathrm{KL}}(p_j^\theta\|q_j)\Big],
\label{eq:ecopd}
\end{equation}
where $\mathcal J(\tau)=\{j\in\{2,\ldots,J\}:|\mathcal A(s_j)|>1\}$ excludes the first action and steps without a choice. Supervision uses student-generated rather than teacher-forced prefixes \citep{agarwal2024gkd}. Only freshly initialized LoRA adapters, parameterized by $\theta$, are trained \citep{hu2022lora}; sampled trajectories and teacher targets are held fixed during each update.

Here, \emph{on-policy} refers to sampling from the current student, not live trading updates. Each rollout is completed and evaluated using labels matured by $t_0$ before teacher targets are computed. The outcome ledger $z(f)$ is supplied only to the teacher.

\begin{table}[t]
\centering
\small
\setlength{\tabcolsep}{4pt}
\renewcommand{\arraystretch}{1.08}
\begin{tabularx}{\columnwidth}{@{}>{\raggedright\arraybackslash}p{0.25\columnwidth}>{\raggedright\arraybackslash}X@{}}
\toprule
Stage & Information and updates\\
\midrule
Student rollout & Context $c$ and generated prefix; no outcome inputs.\\
Teacher targets & Prefix, completed $f$, and ledger $z(f)$; teacher frozen.\\
Distillation & Legal-action targets update student LoRA; labels mature by $t_0$.\\
Generation & Frozen student receives intent, target state, and grammar; no outcome inputs.\\
Screening & Pre-cutoff labels select a new library, then held fixed.\\
Deployment & Lagged evidence updates the pool; matured returns update factor weights.\\
\bottomrule
\end{tabularx}
\caption{Information boundaries. The student receives no outcome inputs and remains frozen during deployment.}
\label{tab:boundaries}
\end{table}

\subsection{Library Regeneration and Screening}
\label{sec:screen}
After training, the student generates a fixed budget of new programs by enumerating economic intents and four target market states (calm or volatile rises and declines; \S\ref{sec:rrg}), without reference programs or outcome feedback as inputs. On pre-cutoff data, for each intent-allowed horizon in $\mathcal H$, screening regresses dailyized top-$K$ returns above the screen-eligible stock mean on six style-return streams and subtracts their fitted terms, retaining the intercept. It tests both signs of the adjusted series, requiring positive means in both chronological halves. Each candidate's best passing horizon/sign pair is ranked by mean-to-standard-deviation ratio; a per-intent cap limits selection into $\mathcal F$. The chosen sign orients the deployed factor. This signed-series proxy does not re-evaluate the oppositely ranked portfolio (Appendix~\ref{app:impl}).

\subsection{RRG: Regime-Residual Gate}
\label{sec:rrg}
RRG adapts factor weights using matured performance, shrinking sparse state-specific estimates toward overall means \citep{demiguel2009optimal}. It adds a bounded deviation to equal weighting (\textbf{Ch2}).

\paragraph{Deployment ledger and market state.} For signal-close date $u$, $\rho_{f,u}$ is factor $f$'s top-$K$ return above the eligible pool mean from close $u$ to $u+h_{\mathrm g}$, where $h_{\mathrm g}$ is the ledger horizon. At trading date $t$, only records with $u+h_{\mathrm g}+1\le t-1$ are readable. All offsets count trading sessions. State $g_t$ uses observations through $t-1$: the sign of the sum of equal-weighted market returns over $L_1$ sessions and whether their $L_1$-session volatility exceeds their $L_2$-session volatility, with $L_1<L_2$. These define calm or volatile rises and declines.

\paragraph{Shrunk state-conditional estimate.} Among the last $W$ readable records, let $n_t$ count those whose signal-close state matches $g_t$, $\bar\rho_f^{\,g_t}$ their mean, and $\bar\rho_f$ the mean of all records. Shrinkage strength $\kappa>0$ gives
\begin{equation}
\hat\mu_{f,t}=\frac{n_t\bar\rho_f^{\,g_t}+\kappa\bar\rho_f}{n_t+\kappa},
\label{eq:shrink}
\end{equation}
using $\bar\rho_f$ when $n_t=0$.

\paragraph{Equal-weight-anchored allocation.} Let $\zeta_{f,t}$ standardize $\hat\mu_{f,t}$ across factors to zero mean and unit sample standard deviation. With mixing strength $\alpha\in[0,1]$ and temperature $\eta>0$,
\begin{equation}
w_{f,t}=\frac{1-\alpha}{|\mathcal F|}+\alpha\,\mathrm{softmax}_{f\in\mathcal F}(\zeta_{f,t}/\eta).
\label{eq:rrg}
\end{equation}
Every factor retains at least $(1-\alpha)/|\mathcal F|$, and $\alpha=0$ recovers equal weighting. Insufficient records or zero dispersion also give equal weights; thresholds and all numerical settings are in Table~\ref{tab:hyper}.

\section{Experiments}
\label{sec:exp}

\begin{table*}[t]
\centering
\small
\setlength{\tabcolsep}{5pt}
\begin{tabular}{l rrrr rrr}
\toprule
& \multicolumn{4}{c}{Factor quality} & \multicolumn{3}{c}{Portfolio}\\
\cmidrule(lr){2-5}\cmidrule(lr){6-8}
Method & IC (\%) & ICIR & RIC (\%) & RICIR & ARR (\%) & SR & MDD (\%)\\
\midrule
\multicolumn{8}{l}{\emph{Market indices}}\\
CSI300 & -- & -- & -- & -- & 10.7 & 0.68 & \textbf{10.5}\\
CSI500 & -- & -- & -- & -- & 19.4 & 0.86 & 19.1\\
CSI1000 & -- & -- & -- & -- & 12.0 & 0.58 & 22.9\\
ChiNext & -- & -- & -- & -- & 34.2 & 1.07 & 25.8\\
STAR50 & -- & -- & -- & -- & 39.4 & 1.08 & 28.1\\
\midrule
\multicolumn{8}{l}{\emph{Formulaic and RL factor mining}}\\
Alpha101 & 1.30 & 0.11 & 0.16 & 0.01 & 20.3 & 1.11 & \underline{10.7}\\
AlphaGen & 3.94 & 0.15 & 1.75 & 0.07 & 15.2 & 0.98 & 15.1\\
AlphaSAGE & 3.77 & 0.24 & 1.56 & 0.08 & 15.0 & 0.89 & 12.9\\
\midrule
\multicolumn{8}{l}{\emph{LLM-agentic factor mining}}\\
RD-Agent & \underline{4.63} & 0.31 & 1.59 & 0.09 & 14.4 & 0.91 & 18.9\\
AlphaAgent & 1.90 & 0.12 & 1.00 & 0.06 & 15.1 & 0.94 & 15.3\\
AlphaMemo & 1.96 & 0.10 & 3.08 & 0.16 & 14.4 & 1.00 & 13.2\\
AlphaBench-EA & 4.34 & 0.17 & 2.21 & 0.09 & 13.7 & 0.94 & 13.0\\
QuantaAlpha & 1.96 & 0.11 & 3.79 & 0.21 & 16.2 & 1.03 & 16.0\\
\midrule
\multicolumn{8}{l}{\emph{LLM post-training}}\\
QuantEvolver & 3.48 & \underline{0.37} & 1.65 & 0.16 & 11.4 & 0.68 & 18.1\\
\midrule
\multicolumn{8}{l}{\emph{Ours}}\\
AlphaLedger (CSI500) & 3.73 & 0.35 & \underline{5.29} & \underline{0.43} & \underline{31.6} & \underline{1.43} & 12.3\\
\rowcolor{oursrow}
AlphaLedger (EviPool) & \textbf{4.98} & \textbf{0.47} & \textbf{7.22} & \textbf{0.62} & \textbf{55.4} & \textbf{2.06} & 10.8\\
\bottomrule
\end{tabular}
\caption{Main results on the test period (January 2, 2025 to July 31, 2026). All strategy rows except AlphaLedger (EviPool) trade point-in-time CSI500 constituents under the same account; indices are price series without trading costs. Shading identifies the full model. Lower MDD is better. \textbf{Bold}: best in this table; \underline{underline}: best among methods trading CSI500. Additional baselines are in Appendix Table~\ref{tab:mlbase}.}
\label{tab:main}
\end{table*}

\subsection{Dataset and Metrics}
\label{sec:data}
We study the Chinese A-share market using daily price, volume, and money-flow data from Tushare\footnote{\url{https://tushare.pro}} and issuer-linked news with sentiment scores from Cailian Press (CLS)\footnote{\url{https://www.cls.cn}}. Our experiments use the news and money-flow channels. The development period (February 2021 to December 2024) is used for all training and selection, and every label it uses ends by December 31, 2024; the test period (January 2, 2025 to July 31, 2026) is used only for backtesting. We report IC, ICIR, rank IC (RIC), and RICIR against 20-day forward returns, and annualized net return (ARR), Sharpe ratio (SR), and maximum drawdown (MDD) after transaction costs (Appendix~\ref{app:metrics}).

\begin{figure*}[t]
\centering
\includegraphics[width=\textwidth]{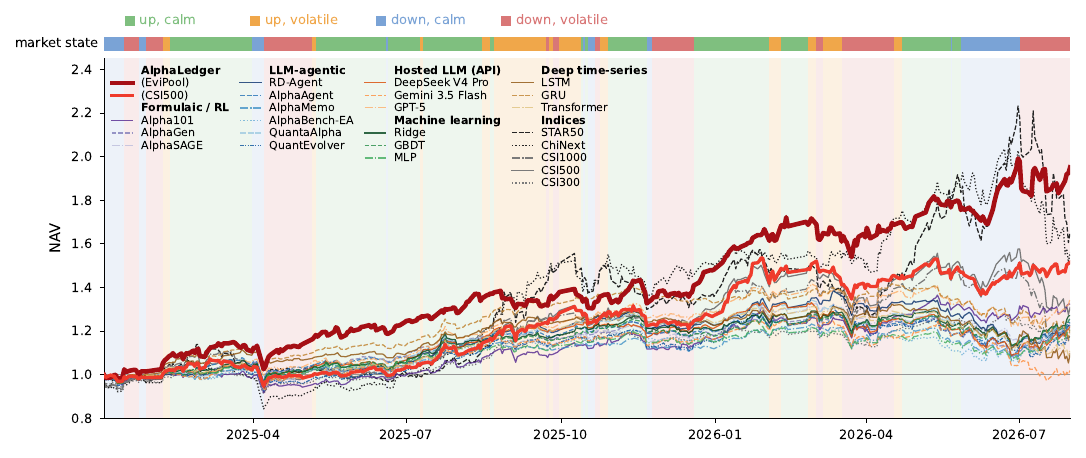}
\caption{Net asset value (NAV) after 5/15 bps trading costs from January 2, 2025, to July 31, 2026. Bold curves show AlphaLedger in EviPool and CSI500. The CSI500 variant and all baselines share a common CSI500 constituent universe for a like-for-like comparison.}
\label{fig:nav}
\end{figure*}

\subsection{Experimental Setup}
\label{sec:setup}
\paragraph{Trading protocol.} Given a universe $\mathcal U_t$, every method is evaluated with the same long-only account. (1) \textbf{Signal generation.} After the close of day $t-1$, the method scores every stock in $\mathcal U_t$. (2) \textbf{Entry rule.} Every five trading days, the account targets the top 40 stocks of $\mathcal U_t$ with equal weights, executing at the close of day $t$, one day after the signal. (3) \textbf{Exit rule.} At each rebalance, at most ten of the lowest-ranked holdings are replaced by the highest-ranked non-holdings; other positions are held.
(4) \textbf{Execution.} Suspended and limit-locked stocks cannot be traded, idle capital stays in cash, and trades pay 5 bps to buy and 15 bps to sell.

\paragraph{Models and baselines.} The student is Qwen3-14B-Base and the teacher is Qwen3-30B-A3B-Base \citep{yang2025qwen3}. All strategy baselines trade point-in-time CSI500 constituents; the full framework uses EviPool. Market indices are separate price-series benchmarks. Baselines span seven families: market indices; formulaic and RL factor mining (Alpha101, AlphaGen, AlphaSAGE); LLM-agentic factor mining (RD-Agent, AlphaAgent, AlphaMemo, AlphaBench-EA, QuantaAlpha); LLM post-training (QuantEvolver); hosted LLMs that write factors directly (DeepSeek V4 Pro, Gemini 3.5 Flash, GPT-5); machine learning on tabular and Qlib features \citep{yang2020qlib}; and deep time-series models (LSTM, GRU, Transformer)
(Appendix~\ref{app:mlbase}).

\begin{figure*}[t]
\centering
\includegraphics[width=0.86\textwidth]{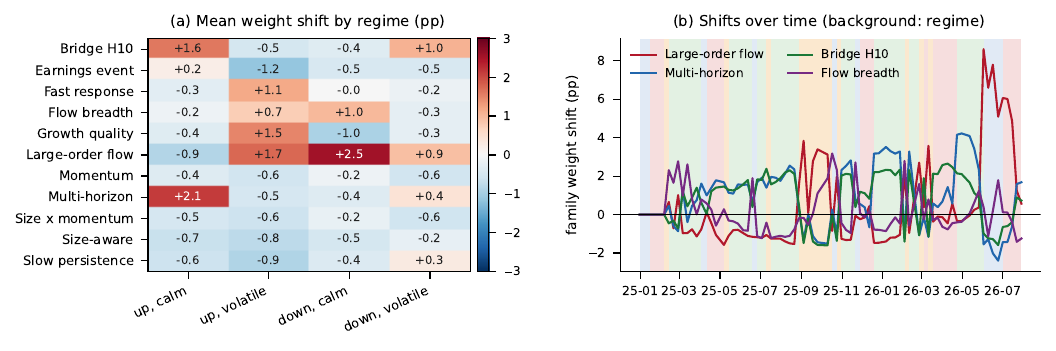}
\caption{RRG weight shifts relative to equal weights: (a) mean shift of each intent family by market state; (b) shifts over time for the four families with the largest tilts.}
\label{fig:regime}
\end{figure*}

\subsection{Main Results}
\label{sec:main}

\paragraph{Overall comparison.}
% Table~\ref{tab:main} reports indices and factor-mining baselines; the remaining baselines are in Appendix Table~\ref{tab:mlbase}. In the common CSI500 universe, AlphaLedger earns 31.6\% a year with a Sharpe ratio of 1.43, against 22.0\% and 1.26 for the strongest baseline (GPT-5), 20.3\% for Alpha101, which has the smallest drawdown (10.7\%), and 13.3\% on average over the 22 strategy baselines. It also attains the highest RIC (5.29\%) and RICIR (0.43) among CSI500 methods. In EviPool, the full framework raises RIC to 7.22\% and earns 55.4\% a year (Sharpe 2.06, drawdown 10.8\%), whereas ChiNext and STAR50 earn 34--39\% with drawdowns of 26--28\%.

Table~\ref{tab:main} reports market indices and factor-mining baselines, with additional baselines in Appendix Table~\ref{tab:mlbase}.
Within the common CSI500 universe, AlphaLedger achieves an annualized return of 31.6\% and a Sharpe ratio of 1.43, compared with 22.0\% and 1.26 for GPT-5, the strongest strategy baseline by annualized return.
It also attains the highest RIC (5.29\%) and RICIR (0.43) among CSI500 methods.
Alpha101 has a lower maximum drawdown (10.7\% versus 12.3\%), but a lower annualized return of 20.3\%.
With EviPool, the full framework achieves a 55.4\% annualized return, a Sharpe ratio of 2.06, and a maximum drawdown of 10.8\%, with RIC reaching 7.22\%.
For context, ChiNext and STAR50 return 34.2\% and 39.4\% annually, respectively, with larger maximum drawdowns of 25.8\% and 28.1\%.

\paragraph{NAV trajectories.}
Figure~\ref{fig:nav} places these results across contrasting market conditions, including the April 2025 tariff-related selloff and the July 2026 technology-stock correction. The CSI500 accounts move closely through 2025; AlphaLedger pulls ahead in early 2026 (1.46 by May vs.\ at most 1.33 for any baseline) and loses only 6.1\% in the June--July decline, against 9.0\% for GPT-5 and 13.5\% for the GRU. Together with the full-period drawdown statistics, these trajectories show that AlphaLedger combines higher terminal wealth with smaller peak-to-trough losses than the two growth-oriented indices over this test period.

\begin{table}[t]
\centering
\small
\setlength{\tabcolsep}{3pt}
\begin{tabular}{lrrrr}
\toprule
Universe & Size & NAV & SR & MDD (\%)\\
\midrule
CSI300 & 300 & 1.304 & 1.06 & 7.9\\
CSI500 & 500 & 1.516 & 1.43 & 12.3\\
CSI800 & 800 & 1.251 & 1.03 & 14.5\\
CSI1000 & 1,000 & 1.429 & 1.49 & 14.5\\
ChiNext & 100 & 1.540 & 1.15 & 17.8\\
STAR50 & 50 & 1.740 & 1.56 & 23.4\\
\midrule
EviPool: news & 500 & 1.525 & 1.48 & 12.0\\
EviPool: flow & 500 & 1.700 & 1.77 & 14.7\\
\rowcolor{oursrow}
\textbf{EviPool (full)} & 500 & \textbf{1.951} & \textbf{2.06} & 10.8\\
\bottomrule
\end{tabular}
\caption{Universe ablation. Universe and evidence-channel comparisons
with the trained generator and shared trading account. Shading identifies the full configuration; lower MDD is better.}
\label{tab:universe}
\end{table}

\subsection{Robustness}
\label{sec:robust}

We evaluate five libraries sampled from the same student checkpoint under identical deployment settings (Appendix Table~\ref{tab:seeds}). Mean terminal NAV is 1.906, and even the lowest-performing seed reaches 1.81. Against the strong growth-index performance in Table~\ref{tab:main}, mean SR is 1.868, 73.0\% above STAR50 (1.08), while mean MDD is 15.3 percentage points lower (12.76\% vs.\ 28.1\%). Every seed exceeds both ChiNext and STAR50 in SR, although MDD varies from 10.34\% to 17.30\%. 

\subsection{Ablations}
\label{sec:ablation}

\paragraph{Universe.}
Table~\ref{tab:universe} compares deployment across universes. EviPool achieves NAV 1.951 and SR 2.06, versus NAV 1.251--1.740 and SR 1.03--1.56 across the listed index universes. Relative to CSI500, its cumulative net return is 43.5 percentage points higher, with MDD reduced from 12.3\% to 10.8\%. CSI300 has the lowest drawdown (7.9\%), but a lower terminal NAV of 1.304. 

\begin{table}[t]
\centering
\small
\setlength{\tabcolsep}{3pt}
\begin{tabular}{lrrrr}
\toprule
Variant & NAV & ARR & SR & MDD\\
\midrule
\multicolumn{5}{l}{\emph{Module removal}}\\
w/o EviPool (CSI500) & 1.516 & 31.6 & 1.43 & 12.3\\
w/o EC-OPD (Student-Direct)  & 1.596 & 36.1 & 1.36 & 13.5\\
w/o RRG (equal weight) & 1.689 & 41.3 & 1.77 & 11.1\\
\midrule
\multicolumn{5}{l}{\emph{Replace EC-OPD with}}\\
EC-OPD $\to$ SFT & 1.527 & 32.2 & 1.09 & 25.7\\
EC-OPD $\to$ OPD & 1.603 & 36.5 & 1.54 & 12.5\\
EC-OPD $\to$ Teahcer-Direct & 1.620 & 37.4 & 1.55 & 18.4\\
EC-OPD $\to$ PPO & 1.688 & 41.3 & 1.34 & 18.5\\
EC-OPD $\to$ GRPO & 1.792 & 46.9 & 1.46 & 17.1\\
\midrule
\rowcolor{oursrow}
\textbf{AlphaLedger} & \textbf{1.951} & \textbf{55.4} & \textbf{2.06} & \textbf{10.8}\\
\bottomrule
\end{tabular}
\caption{Component ablation (ARR, MDD in \%). Training variants share the generation budget, EviPool, and RRG; alternative objectives replace EC-OPD. Shading identifies the full model; lower MDD is better.}
\label{tab:component}
\end{table}

\paragraph{Components.}
Table~\ref{tab:component} shows that removing any component lowers both return and risk-adjusted performance.
Among module removals, excluding EviPool causes the largest ARR decline (55.4\% to 31.6\%), whereas excluding EC-OPD causes the largest SR decline (2.06 to 1.36).
EC-OPD outperforms standard OPD, direct teacher generation, and SFT/PPO/GRPO across all reported metrics.
Compared with GRPO, the highest-return alternative, it improves ARR by 8.5 percentage points and SR by 0.60, while reducing MDD from 17.1\% to 10.8\%.
Replacing RRG with equal weighting lowers ARR to 41.3\% and SR to 1.77, supporting the benefit of adaptive factor allocation in this evaluation.
% \paragraph{Components.}
% Table~\ref{tab:component} shows that removing any component reduces both return and risk-adjusted performance. Removing EviPool causes the largest decline in ARR (55.4\% to 31.6\%), while removing EC-OPD causes the largest decline in SR (2.06 to 1.36). EC-OPD also outperforms SFT, PPO, and GRPO on all reported metrics. Compared with GRPO, the strongest alternative objective, it improves ARR by 8.5 percentage points and SR by 0.60, while reducing drawdown magnitude from 17.1\% to 10.8\%. Replacing RRG with equal weighting lowers ARR to 41.3\% and SR to 1.77, supporting the benefit of adaptive factor allocation in this evaluation.

\subsection{Case Studies}
\label{sec:analysis}
Figure~\ref{fig:regime} groups the 40 library factors by the eleven economic intents the student was prompted with (e.g., large-order flow, multi-horizon consensus, earnings events) and reports each family's total RRG weight minus its equal-weight share. In panel (a), calm rises favor multi-horizon consensus and ten-day bridging factors, volatile rises favor growth quality, fast response, and large-order flow, and calm declines give large-order flow the largest tilt (+2.5 points). Large-order-flow factors pick stocks at the 88th percentile of market value (28th for the equal-weight library), so this allocation favors larger stocks with positive large-order-flow signals. Panel (b) shows this tilt in the June 2026 sell-off: shrinkage lets RRG raise large-order flow to +8.6 points above its equal-weight share before the new state accumulates evidence, coinciding with the account's recovery to 1.951, and the tilt recedes to +0.6 as newly matured records favor other families.

\paragraph{Characteristics of generated factors.}
The trained student generates compact programs: 435 of 437 candidates are distinct, with an average length of approximately 14 grammar tokens. Collectively, these candidates use 43 input fields covering price and volume, money flow, valuation, and news. All 40 selected factors combine at least two input categories and . The selected set also retains coverage of all eleven prompted intents (Appendix~\ref{app:factors}).
% \paragraph{What the student writes.}
% The trained student writes compact, multi-source programs: 435 of its 437 candidates are distinct, and they draw on 43 input fields spanning price and volume, money flow, valuation, and news within about 14 grammar tokens each. Screening sharpens this toward cross-source evidence: every one of the 40 selected factors combines at least two data sources, and all eleven intents remain represented, providing multiple intent families for RRG to reweight (Appendix~\ref{app:factors}).

\paragraph{Style sensitivity.}
To assess dependence on familiar risk exposures, we project rebalance targets onto portfolios that match the eligible EviPool mean on five reconstructed CNE5 technical styles (Appendix~\ref{app:case}). The joint control reaches NAV 1.388 and SR 1.30, versus NAV 1.865 for the coverage-matched reference. It still finishes above CSI500 and CSI1000, but the reduction shows that exposure constraints materially change performance. This is a sensitivity analysis of the selected holdings, not a causal decomposition: the projection changes stock selection and weights, and the descriptors omit industry and fundamental risks.

\section{Conclusion}\label{sec:concl}
We presented AlphaLedger, which trains a local factor generator using portfolio-grounded distillation and deploys it with an evidence-driven universe and regime-aware weights. EC-OPD supplies teacher guidance on student-generated trajectories; EviPool and RRG adapt eligibility and allocation using lagged evidence. Experiments show competitive performance in EviPool and CSI500, with component ablations and repeated-generation evaluation. Future work will explore generator updates as evidence matures, cross-market transfer, and additional multimodal sources.\label{concl:end}

\section*{Limitations}\label{sec:limits}
\paragraph{Fixed generator and limited replication.}
The student is trained once with a frozen teacher, and the screened factor library remains fixed during testing; only the universe and factor weights adapt. Five generation seeds assess sampling variability from one checkpoint, not training-run uncertainty or statistical significance across market periods.

\paragraph{Evaluation comparability.}
Sharing a universe and trading account does not equalize all methods' candidate budgets or aggregation rules. In particular, hosted-LLM baselines generate 60 candidates versus 440 student generation slots. Their results compare the reported configurations rather than equal-compute performance. Pretraining overlap with the historical test period cannot be ruled out for models released after it began.

\paragraph{Market and trading scope.}
Experiments cover Chinese A-shares under a long-only protocol. Transfer to other markets and long--short trading remain untested. The style analysis includes five reconstructed technical descriptors; its unexplained return is not a complete risk-adjusted alpha estimate. Extending the evaluation would require additional markets, risk controls, and short-borrow availability and costs.

\section*{Ethics Statement}
The system produces research signals, not investment advice. Running the generator locally keeps proprietary data in-house. Backtests do not guarantee future returns, and trading on these signals carries financial risk.

\bibliography{custom}

\clearpage
\appendix

\section{Evaluation Metrics}
\label{app:metrics}
Let $V_0=1$ and $V_\tau$ be the account NAV after costs at the close of day $\tau=1,\dots,T$, with daily return $r_\tau=V_\tau/V_{\tau-1}-1$. \emph{Terminal NAV} is $V_T$, and \emph{stress NAV} is $V_T$ with costs doubled to 10 bps to buy and 30 bps to sell. \emph{Annualized return} (ARR) is $V_T^{252/T}-1$. The \emph{Sharpe ratio} (SR) is $\sqrt{252}\,\bar r/\mathrm{sd}(r)$ with a zero risk-free rate. \emph{Maximum drawdown} (MDD) is reported as the positive magnitude $\max_\tau\big(1-V_\tau/\max_{0\le u\le\tau}V_u\big)$; lower is better. \emph{Days above an index} is the fraction of days on which $V_\tau$ exceeds the index level rebased to 1 at the start of the period. For factor metrics, let $s_{i,d}$ be a method's final score for stock $i$ on rebalance date $d$ and $y_{i,d}$ the stock's return from the close of $d+1$ to the close of $d+21$, both restricted to the universe the method trades. The \emph{IC} on date $d$ is the Pearson correlation of $s$ and $y$ after clipping each at its 1st and 99th cross-sectional percentiles, and the \emph{rank IC} (RIC) is the Spearman correlation. We summarize each series by its absolute mean $|\mathbb E[\mathrm{IC}_d]|$ and report ICIR as $|\mathbb E[\mathrm{IC}_d]|/\mathrm{sd}(\mathrm{IC}_d)$ over the rebalance dates, and likewise for RIC. Style exposures are cross-sectional z-scores of the holdings relative to CSI1000 constituents on the same date.

\section{Baseline Details}
\label{app:mlbase}
Table~\ref{tab:mlbase} reports the machine-learning, deep time-series, and hosted-LLM baselines under the protocol of Table~\ref{tab:main}. All baselines are selected or trained once with 20-day labels that mature before 2025 and are then frozen.
\emph{Formulaic and RL factor mining}: the 30 Alpha101-style formulas, the final AlphaGen pool with its learned weights, and the AlphaSAGE library.
\emph{LLM-agentic factor mining and LLM post-training}: the factor libraries of RD-Agent, AlphaAgent, AlphaMemo, AlphaBench-EA, QuantaAlpha, and QuantEvolver, produced with each method's released code on our data. For these library baselines, we orient each factor by its 20-day RIC on CSI500 from July 2023 to December 2024 and keep up to 40 factors with the largest absolute RIC, combined with equal weights; AlphaGen keeps its own weights.
\emph{Machine learning}: Ridge, gradient-boosted trees (GBDT), and a two-layer MLP on 36 price, flow, valuation, and news features, and LightGBM and MLP on the Qlib Alpha158 and Alpha360 feature sets \citep{yang2020qlib}.
\emph{Deep time-series models}: LSTM, GRU, and Transformer encoders on Alpha360 sequences (60 days $\times$ 6 fields), trained for a fixed ten epochs without test-time model selection. The machine-learning and deep models are trained on CSI500 constituents from January 2022 to December 2024.

\begin{table*}[t]
\centering
\small
\setlength{\tabcolsep}{5pt}
\begin{tabular}{l rrrr rrr}
\toprule
& \multicolumn{4}{c}{Factor quality} & \multicolumn{3}{c}{Portfolio}\\
\cmidrule(lr){2-5}\cmidrule(lr){6-8}
Method & IC (\%) & ICIR & RIC (\%) & RICIR & ARR (\%) & SR & MDD (\%)\\
\midrule
\multicolumn{8}{l}{\emph{Machine learning}}\\
Ridge & 2.87 & 0.11 & 3.15 & 0.12 & 17.2 & 1.06 & 12.7\\
GBDT & 2.21 & 0.10 & 2.38 & 0.10 & 16.5 & 1.06 & \underline{11.7}\\
MLP & 2.46 & 0.11 & 2.59 & 0.11 & 13.0 & 0.87 & 12.8\\
LightGBM (Alpha158) & 3.26 & 0.15 & 1.74 & 0.08 & 11.2 & 0.72 & 13.9\\
MLP (Alpha158) & 0.92 & 0.06 & 2.44 & 0.15 & 15.9 & 0.95 & 12.6\\
LightGBM (Alpha360) & 1.89 & 0.10 & 2.73 & 0.13 & 7.6 & 0.53 & 17.7\\
MLP (Alpha360) & 3.23 & 0.27 & 0.53 & 0.04 & 10.8 & 0.64 & 14.8\\
\midrule
\multicolumn{8}{l}{\emph{Deep learning}}\\
LSTM & 2.20 & 0.11 & 3.76 & 0.18 & 6.3 & 0.43 & 20.4\\
GRU & 1.31 & 0.07 & 4.07 & 0.20 & 13.4 & 0.81 & 18.3\\
Transformer & 1.53 & 0.07 & 4.24 & 0.20 & 7.3 & 0.52 & 17.1\\
\midrule
\multicolumn{8}{l}{\emph{Hosted LLMs (API-Direct)}}\\
DeepSeek V4 Pro & 0.84 & 0.05 & 3.33 & 0.20 & 14.1 & 0.87 & 18.2\\
Gemini 3.5 Flash & \underline{4.52} & 0.33 & 1.42 & 0.09 & 1.3 & 0.16 & 23.3\\
GPT-5 & 2.80 & 0.13 & 2.32 & 0.11 & 22.0 & 1.26 & 13.7\\
\midrule
\multicolumn{8}{l}{\emph{Ours}}\\
AlphaLedger (CSI500) & 3.73 & \underline{0.35} & \underline{5.29} & \underline{0.43} & \underline{31.6} & \underline{1.43} & 12.3\\
\textbf{AlphaLedger (EviPool)} & \textbf{4.98} & \textbf{0.47} & \textbf{7.22} & \textbf{0.62} & \textbf{55.4} & \textbf{2.06} & \textbf{10.8}\\
\bottomrule
\end{tabular}
\caption{Additional machine-learning, deep time-series, and API-Direct baselines under the protocol of Table~\ref{tab:main}. All strategy rows except AlphaLedger (EviPool) trade point-in-time CSI500 constituents. Bold marks the best value in this table; underlining marks the best among CSI500 strategies. Lower MDD is better; Appendix~\ref{app:metrics} defines all metrics.}
\label{tab:mlbase}
\end{table*}

\paragraph{Hosted LLMs.} The API-Direct baselines replace the trained student with a hosted model. Each model receives the grammar, the field list, and an abstract intent, and returns ten expressions per call over six calls (60 candidates, against 440 for our student); it never sees market data, outcomes, or economic feedback. Each library is screened by 2024 RIC on CSI500 (top 40), combined with fixed weights estimated once on pre-2025 CSI500 data, and transferred unchanged to the other universes. In CSI500, GPT-5 is the strongest baseline (22.0\% a year, Sharpe 1.26), while DeepSeek V4 Pro and Gemini 3.5 Flash trail (Table~\ref{tab:mlbase}). Table~\ref{tab:apipools} reports all four universes: AlphaLedger leads in each, by 9.6 points of ARR in CSI500 and by 6 to 46 points elsewhere. EviPool lowers ARR for DeepSeek and GPT-5 relative to CSI500 but raises it for Gemini; AlphaLedger obtains the largest gain.

% \begin{table*}[t]
% \centering
% \small
% \setlength{\tabcolsep}{5pt}
% \begin{tabular}{l rr rr rr rr}
% \toprule
% & \multicolumn{2}{c}{CSI300} & \multicolumn{2}{c}{CSI500} & \multicolumn{2}{c}{CSI1000} & \multicolumn{2}{c}{EviPool}\\
% \cmidrule(lr){2-3}\cmidrule(lr){4-5}\cmidrule(lr){6-7}\cmidrule(lr){8-9}
% Method & ARR (\%) & SR & ARR (\%) & SR & ARR (\%) & SR & ARR (\%) & SR\\
% \midrule
% DeepSeek V4 Pro & $-$1.6 & $-$0.05 & 14.1 & 0.87 & 20.4 & 1.13 & 8.1 & 0.50\\
% Gemini 3.5 Flash & $-$1.0 & 0.00 & 1.3 & 0.16 & $-$1.1 & 0.03 & $-$11.1 & $-$0.47\\
% GPT-5 & 6.2 & 0.50 & 22.0 & 1.26 & 15.2 & 0.84 & 9.0 & 0.54\\
% \midrule
% \textbf{AlphaLedger} & \textbf{19.1} & \textbf{1.06} & \textbf{31.6} & \textbf{1.43} & \textbf{26.6} & \textbf{1.49} & \textbf{55.4} & \textbf{2.06}\\
% \bottomrule
% \end{tabular}
% \caption{API-Direct versus AlphaLedger across four stock universes in the main evaluation period. All accounts share trading rules and costs. API libraries are screened on pre-2025 CSI500 data and transferred unchanged; generation budgets and aggregation rules differ by method. Higher ARR and SR are better; bold marks the best result per column.}
% \label{tab:apipools}
% \end{table*}

\begin{table*}[t]
\centering
\small
\setlength{\tabcolsep}{5pt}
\begin{tabular}{l rr rr rr rr}
\toprule
& \multicolumn{2}{c}{CSI300} & \multicolumn{2}{c}{CSI500} & \multicolumn{2}{c}{CSI1000} & \multicolumn{2}{c}{EviPool}\\
\cmidrule(lr){2-3}\cmidrule(lr){4-5}\cmidrule(lr){6-7}\cmidrule(lr){8-9}
Method & ARR (\%) & SR & ARR (\%) & SR & ARR (\%) & SR & ARR (\%) & SR\\
\midrule
DeepSeek V4 Pro & 1.6 & 0.05 & 14.1 & 0.87 & 20.4 & 1.13 & 8.1 & 0.50\\
Gemini 3.5 Flash & 1.0 & 0.00 & 1.3 & 0.16 & 1.1 & 0.03 & 11.1 & 0.47\\
GPT-5 & 6.2 & 0.50 & 22.0 & 1.26 & 15.2 & 0.84 & 9.0 & 0.54\\
\midrule
\rowcolor{oursrow}
\textbf{AlphaLedger} & \textbf{19.1} & \textbf{1.06} & \textbf{31.6} & \textbf{1.43} & \textbf{26.6} & \textbf{1.49} & \textbf{55.4} & \textbf{2.06}\\
\bottomrule
\end{tabular}
\caption{API-Direct versus AlphaLedger across four stock universes in the main evaluation period. All accounts share trading rules and costs. API libraries are screened on pre-2025 CSI500 data and transferred unchanged; generation budgets and aggregation rules differ by method. Higher ARR and SR are better; bold marks the best result per column.}
\label{tab:apipools}
\end{table*}

\section{Implementation Details}
\label{app:impl}
Table~\ref{tab:hyper} lists all settings; every value was fixed before the evaluations reported here.  The final reported experiments use two NVIDIA H800 GPUs with 80 GB of memory per GPU. 

\begin{table*}[t]
\centering
\small
\setlength{\tabcolsep}{5pt}
\renewcommand{\arraystretch}{1.08}
\begin{tabularx}{\textwidth}{@{}
  >{\raggedright\arraybackslash}p{0.15\textwidth}
  >{\raggedright\arraybackslash}p{0.26\textwidth}
  >{\raggedright\arraybackslash}X
@{}}
\toprule
Component & Setting & Value\\
\midrule
EviPool
& channels, weights $\beta_{\mathrm{news}},\beta_{\mathrm{flow}}$
& news (3-day count, 5-day sentiment), money flow (60-day own history); 0.5, 0.5\\
& size $N$, retention band $M$ & 500, 1{,}000\\
& refresh interval & every 10 pool dates\\
& eligibility $\mathcal E_t$
& price $\ge 3$, listed $\ge 180$ days, 20-day traded value above the 20th percentile\\
\midrule
Economic credit
& reference library $|\mathcal R|$, evaluation portfolio $K$
& 40, 50 (also used in screening and RRG)\\
& horizons $\mathcal H$, cost rate $\gamma$ & $\{5,10,20\}$ days, 0.002\\
& credit signal dates, label cutoff $t_0$
& July--December 2024; December 31, 2024\\
\midrule
EC-OPD
& student
& Qwen3-14B-Base with fresh LoRA (rank 8, scale 16, dropout 0, q/v projections)\\
& teacher & Qwen3-30B-A3B-Base, frozen, no fine-tuning\\
& optimization
& learning rate $2\times10^{-6}$, 468 updates, $\le$4 updates per trajectory\\
& sampling & 11 intents, 12 passes; temperature 0.8, top-$p$ 0.95\\
\midrule
Screening
& candidates
& 440 (2 generation paths $\times$ 220 slots; 11 intents $\times$ 4 market states)\\
& library $|\mathcal F|$, per-intent cap
& 40, 12; development period (Feb.\ 2021--Dec.\ 2024); 6 style exposures removed\\
\midrule
RRG
& residual $\alpha$, temperature $\eta$, shrinkage $\kappa$ & 0.3, 1.0, 10\\
& records $W$, horizon $h_{\mathrm g}$, readable lag & 250, 5 days, 6 days\\
& state windows $L_1$, $L_2$ & 20, 60 days\\
\midrule
Account
& holdings, replacements, rebalance
& top 40, $\le$10 per rebalance, every 5 trading days\\
& costs & 5/15 bps (native)\\
\midrule
ML baselines & Ridge & $\lambda=100$\\
& GBDT & 200 iterations, learning rate 0.05, 31 leaves\\
& MLP & hidden units 128--64, 15 epochs\\
\bottomrule
\end{tabularx}
\caption{Hyperparameters and settings of AlphaLedger and the machine-learning baselines.}
\label{tab:hyper}
\end{table*}

\section{Factor Grammar, Intents, and Example Factors}
\label{app:factors}
\paragraph{Grammar.} Factors are programs in a typed grammar whose leaves are input fields from five data families: price (P; e.g., open, daily return), liquidity (L; average traded value over 5/10/20 days, free-float turnover), money flow (F; net inflow, large-order imbalance, extra-large-order ratio, reversal flow), valuation and size (V; PE, PB, PS, market value), and issuer-linked news (N; item counts, sentiment ratios, and topic counts such as policy, management, and macro). Time-series operators (\texttt{ewma}, \texttt{ts\_mean}, \texttt{ts\_std}, \texttt{delta}, \texttt{delay}) take windows from $\{3,5,10,20,60\}$ days; cross-sectional operators (\texttt{rank}, \texttt{zscore}, \texttt{demean}), arithmetic, and the conditional \texttt{where(greater($\cdot$,0.5),$\cdot$,$\cdot$)} combine them. Every program is evaluated on data up to the previous close.

%\paragraph{Intents.} The generation context $c$ states one of eleven economic intents (Table~\ref{tab:intents}). An intent specifies what the factor should express and, for some intents, which data family it must use and at which horizons it may be evaluated; it does not prescribe a formula, a sign, or a later portfolio weight. Generation slots are interleaved so that every intent receives the same number of proposals. The intents label the families analyzed in Figure~\ref{fig:regime}. %wzk
\paragraph{Intents.}
The generation context $c$ states one of eleven economic intents
(Table~\ref{tab:intents}). An intent describes the desired signal and,
for some intents, specifies required data families and allowed
evaluation horizons; it does not prescribe a formula, a sign, or a
later portfolio weight. Generation slots are interleaved so that every
intent receives the same number of proposals. Figure~\ref{fig:regime}
groups factors by these generation intents, which describe the
elicitation context rather than independently validated economic
mechanisms. %wzk

\begin{table*}[t]
\centering
\footnotesize
\setlength{\tabcolsep}{4pt}
\begin{tabular}{l p{0.47\textwidth} l c c}
\toprule
Intent & What the factor should express & Required data & Horizons & Count\\
\midrule
Fast response & Short (3--5 day) reaction or repair signal confirmed by another data family & any & 5, 10 & 4\\
Bridge H10 & Signal built on a genuine 10-day window rather than a proxy for 20 days & any & 10 & 3\\
Slow persistence & Persistent 20--60 day signal confirmed by another data family & any & 20 & 4\\
Multi-horizon & Combines evidence from at least two of the short, medium, and long window bands & any & 5, 10, 20 & 7\\
Size-aware & Uses valuation or size to condition or normalize non-size evidence & valuation/size & 5, 10, 20 & 2\\
Size $\times$ momentum & Interaction of size with a price-based style across two window bands & valuation/size, price & 10, 20 & 1\\
Flow breadth & Combines money flow and liquidity participation as a breadth or rotation signal & money flow, liquidity & 5, 10, 20 & 6\\
Momentum & Strength, consistency, or exhaustion of price trends & price & 5, 10, 20 & 1\\
Growth quality & Growth or improvement signals that may conflict with static cheapness & news & 5, 10, 20 & 5\\
Large-order flow & Money-flow participation and order-flow imbalance among actively traded stocks & money flow & 5, 10 & 4\\
Earnings event & Reactions to earnings, policy, and product news, and their decay or drift & news & 5, 10 & 3\\
\bottomrule
\end{tabular}
\caption{The eleven generation intents. ``Required data'' lists data families an intent must use (``any'': no requirement); Horizons: holding periods (days) allowed in screening; Count: number of the 40 selected factors with this intent.}
\label{tab:intents}
\end{table*}

%\paragraph{Example factors.} Table~\ref{tab:examples} shows one selected factor per intent, chosen to illustrate the range of structures in the library. The programs combine evidence across data families through several patterns: additive blends of smoothed ranks (fast response, multi-horizon), multiplicative interactions spanning liquidity, money flow, and news (growth quality), conditional switches that flip a signal depending on another variable (Bridge H10), standardized deviations from a stock's own recent history (slow persistence, large-order flow), and price changes relative to lagged levels (momentum, size $\times$ momentum). Every selected factor reads at least two data families. 
\paragraph{Example factors.}
Table~\ref{tab:examples} shows one selected factor per generation
intent. The examples include additive blends of smoothed ranks,
products of ranked inputs, liquidity-conditioned sign switches,
deviations standardized by a stock's own recent history, and changes
relative to lagged levels. Shared fields and operator patterns recur
across intents, so different intent labels need not imply different
signal behavior. Expressions are shown in their raw generated form,
including algebraically redundant terms; the development-set
orientation is applied before deployment. Each selected factor reads
at least two data families.

\begin{table*}[t]
\centering
\scriptsize
\setlength{\tabcolsep}{4pt}
\begin{tabular}{l c c c p{0.68\textwidth}}
\toprule
Intent & Data & $h$ & Sign & Expression\tabularnewline
\midrule
Fast response & FN & 10 & $-$ & \raggedright\texttt{rank(\allowbreak{}-1 * (\allowbreak{}rank(\allowbreak{}ewma(\allowbreak{}mf\_net\_amount, 20)) + 1 * rank(\allowbreak{}ewma(\allowbreak{}news\_topic\_policy\_3d, 5))))}\tabularnewline
Bridge H10 & LF & 10 & $-$ & \raggedright\texttt{demean(\allowbreak{}where(\allowbreak{}greater(\allowbreak{}rank(\allowbreak{}ewma(\allowbreak{}adv10, 10)), 0.5), rank(\allowbreak{}ewma(\allowbreak{}mf\_reversal\_signal, 10)), -1 * rank(\allowbreak{}ewma(\allowbreak{}mf\_reversal\_signal, 10))))}\tabularnewline
Slow persistence & LN & 20 & $-$ & \raggedright\texttt{rank(\allowbreak{}rank(\allowbreak{}(\allowbreak{}news\_count\_1d - ts\_mean(\allowbreak{}news\_count\_1d, 60)) / (\allowbreak{}ts\_std(\allowbreak{}news\_count\_1d, 60) + 1e-6)) * (\allowbreak{}0.5 * rank(\allowbreak{}(\allowbreak{}adv20 - ts\_mean(\allowbreak{}adv20, 60)) / (\allowbreak{}ts\_std(\allowbreak{}adv20, 60) + 1e-6))))}\tabularnewline
Multi-horizon & LN & 20 & $-$ & \raggedright\texttt{demean(\allowbreak{}rank(\allowbreak{}ewma(\allowbreak{}adv5, 10)) + 0.5 * rank(\allowbreak{}ewma(\allowbreak{}news\_topic\_policy\_3d, 3)) + 0.5 * rank(\allowbreak{}ewma(\allowbreak{}turnover\_rate\_f, 3)))}\tabularnewline
Size-aware & VN & 20 & + & \raggedright\texttt{zscore(\allowbreak{}-1 * (\allowbreak{}rank(\allowbreak{}ewma(\allowbreak{}pb, 10)) - 0.5 * rank(\allowbreak{}ewma(\allowbreak{}news\_topic\_policy\_3d, 10))))}\tabularnewline
Size $\times$ momentum & PV & 10 & $-$ & \raggedright\texttt{zscore(\allowbreak{}-1 * (\allowbreak{}rank(\allowbreak{}ewma(\allowbreak{}pct\_chg, 20)) + 0.5 * rank(\allowbreak{}delta(\allowbreak{}total\_mv, 3) / (\allowbreak{}abs(\allowbreak{}delay(\allowbreak{}total\_mv, 3)) + 1e-6))))}\tabularnewline
Flow breadth & PLF & 20 & $-$ & \raggedright\texttt{demean(\allowbreak{}-1 * (\allowbreak{}rank(\allowbreak{}ewma(\allowbreak{}pct\_chg, 3)) - 0.25 * rank(\allowbreak{}ewma(\allowbreak{}adv5, 3)) - 0.5 * rank(\allowbreak{}ewma(\allowbreak{}mf\_reversal\_signal, 3))))}\tabularnewline
Momentum & PN & 10 & $-$ & \raggedright\texttt{demean(\allowbreak{}-1 * (\allowbreak{}rank(\allowbreak{}delta(\allowbreak{}open, 10) / (\allowbreak{}abs(\allowbreak{}delay(\allowbreak{}open, 10)) + 1e-6)) + 0.25 * rank(\allowbreak{}ts\_std(\allowbreak{}news\_topic\_policy\_3d, 3))))}\tabularnewline
Growth quality & LFN & 20 & $-$ & \raggedright\texttt{demean(\allowbreak{}rank(\allowbreak{}ewma(\allowbreak{}adv10, 3)) * (\allowbreak{}0.5 * rank(\allowbreak{}ewma(\allowbreak{}mf\_reversal\_signal, 3))) * (\allowbreak{}0.25 * rank(\allowbreak{}ewma(\allowbreak{}news\_count\_1d, 3))))}\tabularnewline
Large-order flow & LF & 10 & $-$ & \raggedright\texttt{rank(\allowbreak{}rank(\allowbreak{}(\allowbreak{}adv10 - ts\_mean(\allowbreak{}adv10, 10)) / (\allowbreak{}ts\_std(\allowbreak{}adv10, 10) + 1e-6)) * (\allowbreak{}0.5 * rank(\allowbreak{}ts\_mean(\allowbreak{}mf\_reversal\_signal, 10))))}\tabularnewline
Earnings event & FN & 10 & $-$ & \raggedright\texttt{rank(\allowbreak{}-1 * (\allowbreak{}rank(\allowbreak{}ewma(\allowbreak{}news\_pos\_ratio\_3d, 5)) + 0.5 * rank(\allowbreak{}ewma(\allowbreak{}mf\_reversal\_signal, 5))))}\tabularnewline
\bottomrule
\end{tabular}
\caption{Example factors from the selected library, one per intent, as written by the trained student. Data: data families read (P price, L liquidity, F money flow, V valuation/size, N news); $h$: screening horizon in days; Sign: orientation applied to the expression, set on development labels.}
\label{tab:examples}
\end{table*}

\section{Sampling Robustness}\label{app:seeds}

\begin{figure*}[t]
\centering
\includegraphics[width=\textwidth]{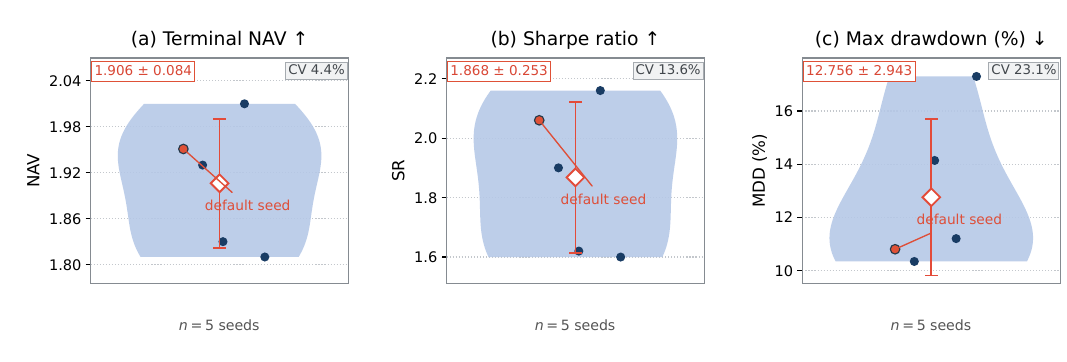}
\caption{Performance across five generation seeds.
Dots show individual seeds, with the default seed used in the main experiments in red.
White diamonds and red whiskers indicate the mean and
$\pm$ one sample standard deviation, respectively.
CV denotes the coefficient of variation.}
\label{fig:seedrobust}
\end{figure*}

\begin{table}[t]
\centering
\small
\setlength{\tabcolsep}{4pt}
\resizebox{\columnwidth}{!}{%
\begin{tabular}{lccc}
\toprule
Seed & NAV $\uparrow$ & SR $\uparrow$ & MDD (\%) $\downarrow$ \\
\midrule
Default & 1.951 & 2.06 & 10.80 \\
s2      & 1.93 & 1.90 & 10.34 \\
s3      & 1.83 & 1.62 & 14.14 \\
s4      & 2.01 & 2.16 & 11.20 \\
s5      & 1.81 & 1.60 & 17.30 \\
\midrule
\rowcolor{oursrow}
Mean $\pm$ SD
& $\mathbf{1.906 \pm 0.084}$
& $\mathbf{1.868 \pm 0.253}$
& $\mathbf{12.76 \pm 2.94}$ \\
\bottomrule
\end{tabular}%
}
\caption{Sampling robustness across five factor libraries generated
from the same student checkpoint and evaluated with EviPool, RRG,
and default transaction costs over January 2, 2025--July 31, 2026.
Bold values report the mean and sample standard deviation across seeds.}
\label{tab:seeds}
\end{table}

Table~\ref{tab:seeds} and Figure~\ref{fig:seedrobust} summarize five generations from one trained checkpoint. Mean terminal NAV is $1.906 \pm 0.084$, SR is $1.868 \pm 0.253$, and MDD is $(12.76 \pm 2.94)\%$ (sample SD). NAV varies less in relative terms (CV: 4.4\%) than MDD (CV: 23.1\%). All five NAVs exceed 1.80, with MDD below 18\%. These repetitions quantify sampling variability; they do not assess independent training runs or additional market periods.

\section{Case Studies}\label{app:case}

\begin{figure*}[t]
\centering
\includegraphics[width=\textwidth]{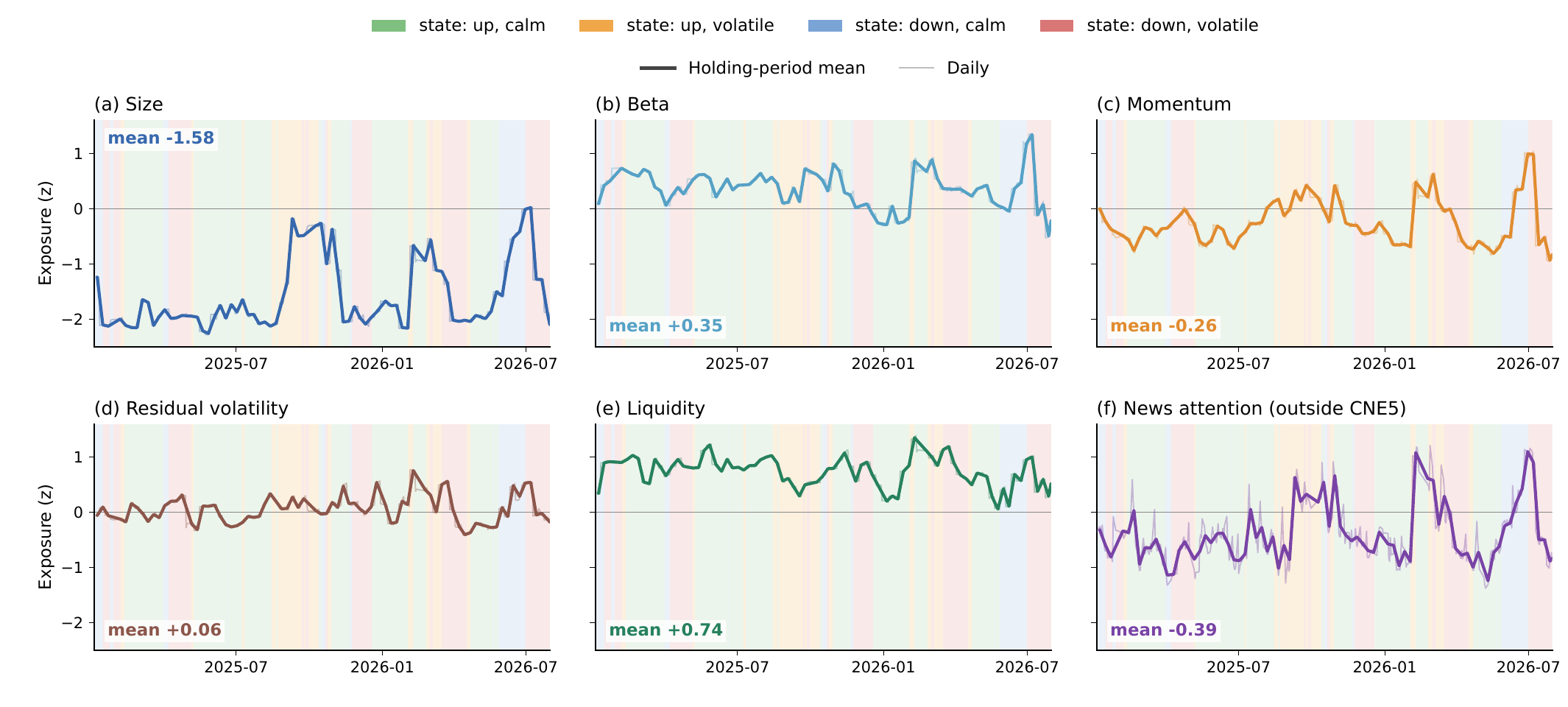}
\caption{Style exposure of the actual holdings, in cross-sectional standard deviations relative to the cap-weighted market mean, on a common scale. (a)--(e) Five CNE5 technical styles; (f) news attention, outside CNE5. Bold lines average each five-day holding period; faint lines are daily values, which step at rebalances and drift between them. Panel labels give time averages; shading marks the market state.}
\label{fig:exposure}
\end{figure*}

\begin{figure*}[t]
\centering
\includegraphics[width=\textwidth]{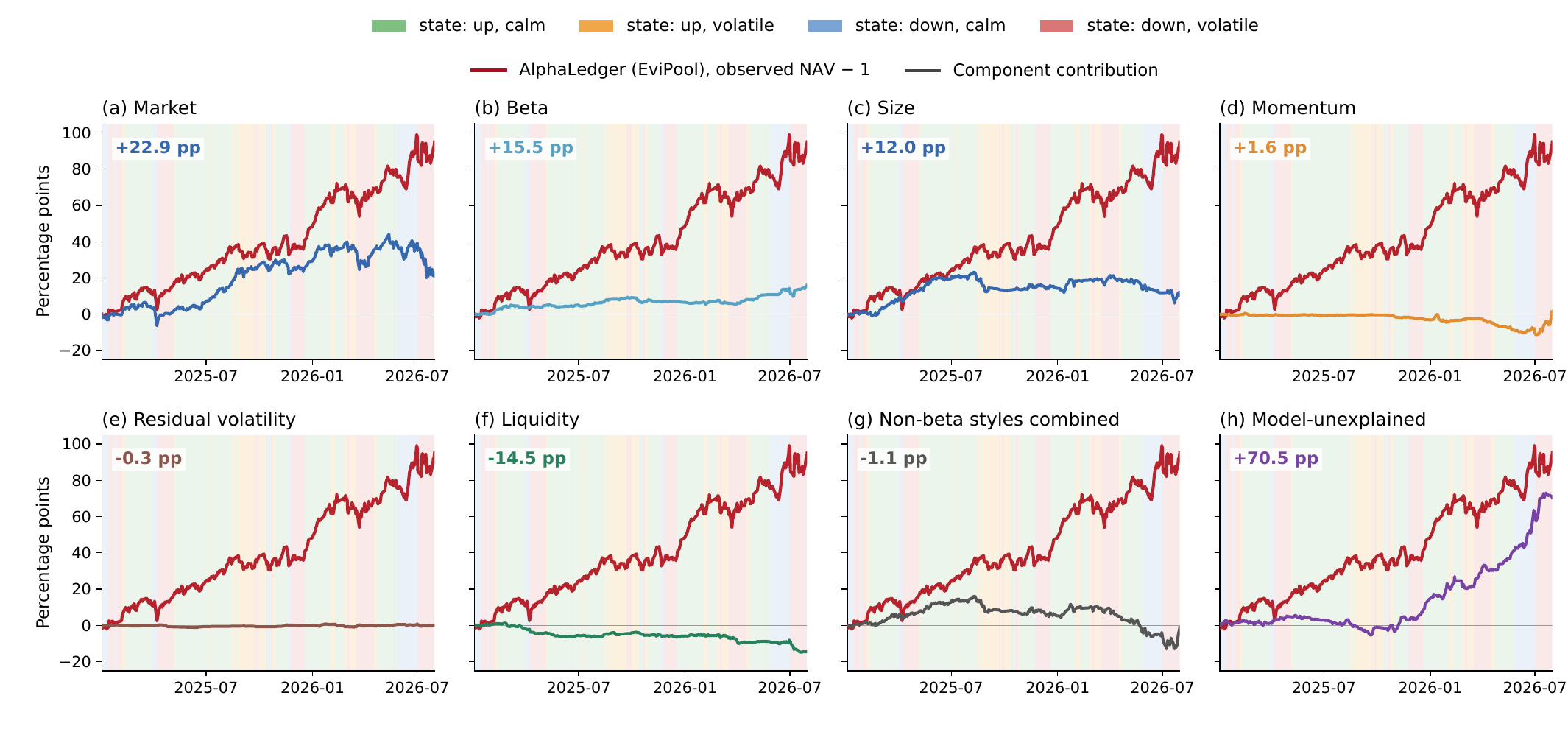}
\caption{Return attribution using a fitted common intercept and five reconstructed CNE5 technical styles; ``Market denotes the common intercept contribution. Each panel plots one component's cumulative contribution against the observed net return of AlphaLedger (EviPool), both in percentage points of initial NAV on a common scale; panel (g) sums size, momentum, residual volatility, and liquidity and is not an additional component. Trading costs are omitted.}
\label{fig:percomp}
\end{figure*}

\begin{figure*}[t]
\centering
\includegraphics[width=\textwidth]{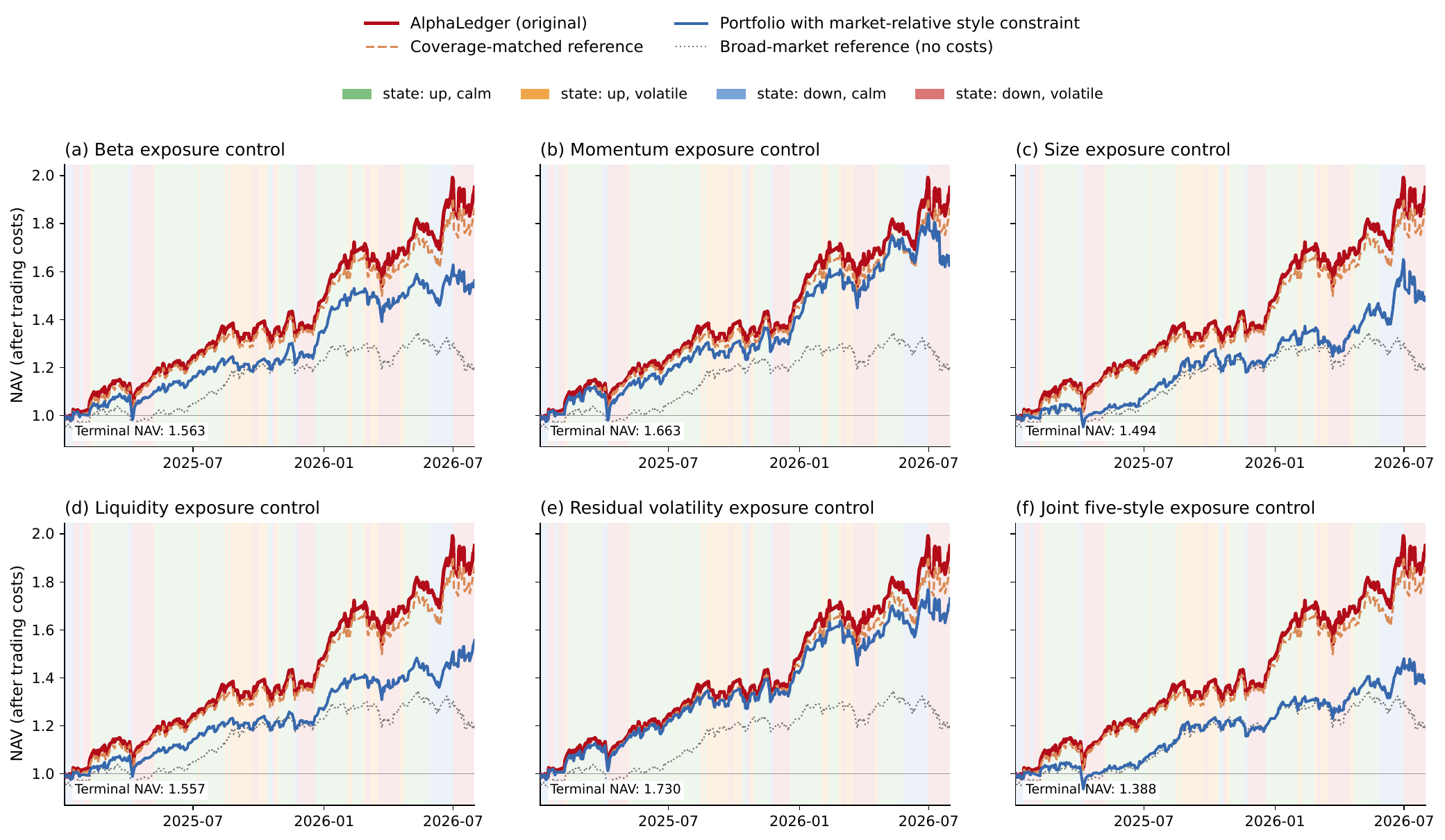}
\caption{NAV of portfolios whose holdings are constrained to the cap-weighted broad-market exposure of each CNE5 technical style, or of all five jointly; candidates remain the eligible EviPool stocks, and the alpha score, RRG weights, and costs are unchanged. Dashed: coverage-matched reference; dotted: cap-weighted broad-market reference without costs.}
\label{fig:controls}
\end{figure*}

This appendix examines the deployed portfolio's technical style exposures (Figure~\ref{fig:exposure}), describes its returns under a specified attribution model (Figure~\ref{fig:percomp}), and re-trades projected portfolios with market-relative style targets (Figure~\ref{fig:controls}). These diagnostics measure sensitivity to five technical styles; they do not identify the learned factors' causal contribution or control all risk exposures.

\paragraph{Style exposures.}
We reconstruct five technical styles from MSCI CNE5 descriptor
definitions: size, beta, momentum, residual volatility, and
liquidity. Descriptors use information through the previous close
and are evaluated on actual beginning-of-day holdings. They are
centered at the cap-weighted market mean and scaled by the
cross-sectional standard deviation; missing held descriptors
contribute zero standardized exposure.
Figure~\ref{fig:exposure} shows average exposures of $-1.58$ for
size, $+0.35$ for beta, $-0.26$ for momentum, $+0.06$ for
residual volatility, and $+0.74$ for liquidity.
Size is the dominant average tilt, but the account temporarily
shifts toward larger stocks around June--July 2026, with size
exposure approaching the market mean before becoming more
negative again. This rotation coincides with the rise and
subsequent retreat of RRG's overweight in large-order-flow
factors (Section~\ref{sec:analysis}).
News attention, shown separately in panel~(f), also rises
temporarily around this rotation despite a full-period mean
of $-0.39$. News-informed stock eligibility thus coexists with
below-market average news attention in the traded holdings:
EviPool determines eligibility, while factor scores and RRG
weights determine the portfolio's subsequent ranking.

\begin{figure*}[t]
\centering
\includegraphics[width=\textwidth]{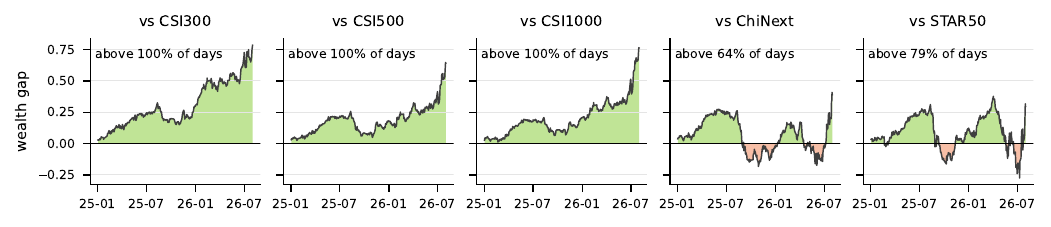}
\caption{NAV of the full framework minus each index over the test period. Green: above the index; red: below.}
\label{fig:gap}
\end{figure*}

\paragraph{Margin over indices.}
AlphaLedger finishes the test period ahead of all five indices,
including ChiNext and STAR50 (Figure~\ref{fig:gap}).
It remains above CSI300, CSI500, and CSI1000 throughout the
displayed period, with substantial positive terminal NAV gaps.
Against ChiNext and STAR50, the account leads on 64\% and 79\%
of trading days, respectively. Although these margins fluctuate,
late-period gains reverse the preceding deficits and establish
positive final margins over both growth indices.
All indices are rebased price series without reinvested
dividends or trading costs.

\begin{figure}[t]
\centering
\includegraphics[width=\columnwidth]{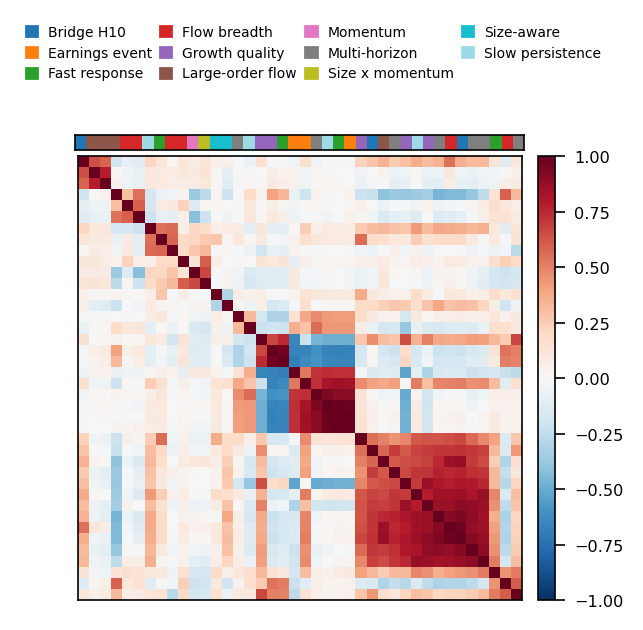}
\caption{Correlation of the 40 selected factors' scores over 2023--2024, ordered by hierarchical clustering. The strip on top marks each factor's intent.}
\label{fig:corr}
\end{figure}

\paragraph{Factor similarity.}
Figure~\ref{fig:corr} shows that the generated factors are neither independent
nor redundant as a whole. Most pairs are weakly correlated (for more than half
of the factors, every pairwise correlation is below $|\rho| = 0.3$ in absolute
value), indicating that the library spans a broad range of distinct behaviors;
the remaining factors form a few coherent groups with moderate-to-strong
internal correlation, including a small number of near-duplicate pairs and a
few factors that move against the dominant group. The groups cut across the
prompting intents: factors obtained under multi-horizon, earnings, growth, and
flow-breadth prompts appear in the same group. We therefore treat the eleven
intents as complementary elicitation strategies that jointly define a factor
space, not as eleven disjoint behaviors, and we report results at both the
intent level and the group level throughout.

% \paragraph{Factor similarity.}
% Figure~\ref{fig:corr} shows both weakly related pairs and groups of strongly correlated factors, including near-duplicates. Correlated groups span several prompting intents. The eleven intents therefore label generation conditions rather than eleven independent behaviors; expression diversity alone does not establish portfolio diversification.

% \paragraph{Return attribution.}
% Figure~\ref{fig:percomp} shows cumulative contributions from a daily cross-sectional regression with an intercept and five reconstructed CNE5 technical styles. The fit uses weighted least squares, with objective weights proportional to the square root of previous-close market capitalization. Each day's fitted contributions are weighted by actual beginning-of-day holdings and linked using beginning NAV. The panel labeled ``Market'' denotes the fitted common intercept contribution, rather than the return on a passive market portfolio. Its contribution is $+22.9$ pp, compared with $+15.5$ pp for beta, $+12.0$ pp for size, $+1.6$ pp for momentum, $-0.3$ pp for residual volatility, and $-14.5$ pp for liquidity. Panel (g) aggregates the four non-beta styles to $-1.1$ pp: their small net contribution masks substantial offsetting components. Panel (h) reports a model-unexplained contribution of $+70.5$ pp. This residual includes stock-specific returns, omitted exposures, and model error; it does not identify the learned factors' causal contribution or provide a formal upper bound on alpha. The plotted component contributions exclude trading costs, whereas the observed return curve is after costs; panel (g) is a subtotal and must not be added to its constituent panels again.

\paragraph{Return attribution.}
Figure~\ref{fig:percomp} decomposes returns using a daily
cross-sectional regression with an intercept and five
reconstructed CNE5 technical styles. The fit uses weighted
least squares, with objective weights proportional to the
square root of previous-close market capitalization.
Contributions are aggregated using actual beginning-of-day
holdings and linked through beginning NAV.
The fitted common intercept, labeled ``Market,'' contributes
$+22.9$ pp, compared with $+15.5$ pp for beta, $+12.0$ pp
for size, $+1.6$ pp for momentum, $-0.3$ pp for residual
volatility, and $-14.5$ pp for liquidity.
The four non-beta styles jointly contribute only $-1.1$ pp
(panel~(g)), as the positive size contribution is largely
offset by liquidity.
The largest contribution is the model-unexplained component
at $+70.5$ pp (panel~(h)), which accumulates strongly in 2026.
Its dominance is consistent with AlphaLedger's learned factors
capturing useful return information beyond the five measured
technical styles.

\paragraph{Style-controlled portfolios.}
We examine whether performance persists when portfolio exposures
are constrained to the cap-weighted broad-market reference used
in Figure~\ref{fig:exposure}.
At each rebalance, the top-40 equal-weight target selected by the
unchanged RRG score is projected onto long-only portfolios of
eligible EviPool stocks, minimizing the $\ell_2$ change while
matching one technical style exposure or all five jointly.
The projection uses previous-close descriptors; factor scores,
RRG weights, and trading costs remain unchanged.

Figure~\ref{fig:controls} shows that single-style controls yield
terminal NAVs from 1.494 for size to 1.730 for residual volatility.
The joint five-style control reaches 1.388, retaining a 38.8\%
cumulative net return. All controlled portfolios finish above
the broad-market reference, although their NAVs are lower than
the coverage-matched reference.
The reduction reflects both the exposure restrictions and the
portfolio changes required to satisfy them. Positive returns
and an advantage over the market reference persist even under
the joint restriction on all five measured styles.

% \paragraph{Style-controlled portfolios.}
% At each rebalance, we project the coverage-matched account's equal-weight target, formed under the same score and retention rule, onto eligible EviPool stocks with complete descriptors. The projection minimizes squared weight changes subject to full investment, long-only weights, a 5\% per-stock cap, and exposure to one style or all five jointly equal to the cap-weighted broad-market reference. Targets use previous-close descriptors, and the account trades with the original execution rules and costs. Figure~\ref{fig:controls} compares the resulting NAV paths. The coverage-matched reference ends at 1.865, versus 1.951 for the original account. Single-style controls end at 1.563 (beta), 1.663 (momentum), 1.494 (size), 1.557 (liquidity), and 1.730 (residual volatility); the joint control ends at 1.388. Every controlled account finishes below the coverage-matched reference and above the plotted broad-market reference. The joint control shows the largest reduction in terminal NAV. These differences measure sensitivity to the complete projection procedure, which changes stock selection and weights as well as style exposures. They do not isolate the return caused by each style. The objective minimizes deviation from the original target; it does not maximize the factor score, and these NAVs are not lower bounds on the returns attainable under the same constraints. Industry and fundamental risks remain uncontrolled.

\begin{figure*}[t]
\centering
\includegraphics[width=\textwidth]{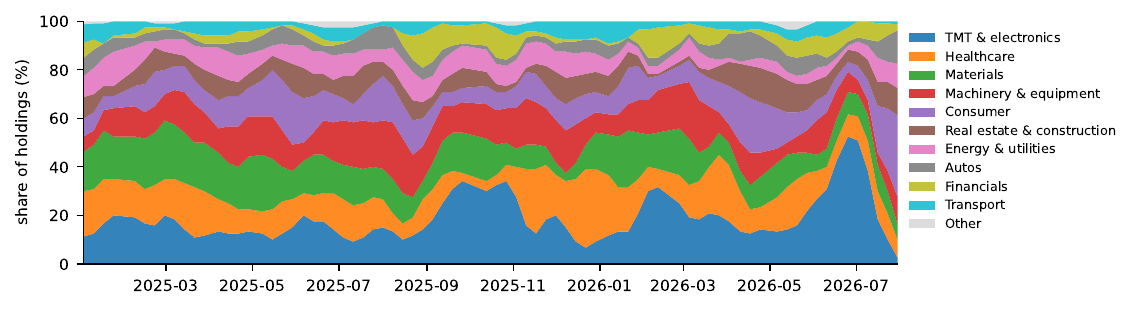}
\caption{Sector composition of the holdings at each rebalance.}
\label{fig:industry}
\end{figure*}

\paragraph{Sector rotation.}
Figure~\ref{fig:industry} shows the holdings rotating without a sector constraint. Materials, the largest sector in early 2025 (20\%), shrink to about 10\% by mid-2026; technology and electronics grow from 16\% to 26\% and become the largest sector, while healthcare stays between 12\% and 19\%. The rotation follows the market's leadership over the period and emerges from the factor scores and the moving universe.

\end{document}